\pdfoutput=1
\documentclass[twocolumn,english,aps,superscriptaddress,prb]{revtex4-1}

\usepackage{babel}
\usepackage{amsmath}
\usepackage{amssymb}
\usepackage{graphicx}

\begin{document}

\title{Dimerization and effective decoupling in two spin-1 generalizations of the spin-1/2 Majumdar-Ghosh chain}

\author{Natalia Chepiga}
\affiliation{Institute for Theoretical Physics, University of Amsterdam, Science Park 904 Postbus 94485, 1090 GL Amsterdam, The Netherlands}
\author{Fr\'ed\'eric Mila}
\affiliation{Institute of Theoretical Physics, Ecole Polytechnique F\'ed\'erale de Lausanne (EPFL), CH-1015 Lausanne, Switzerland}

\date{\today}
\begin{abstract} 
We perform a systematic DMRG investigation of the two natural spin-1 generalizations of the spin-1/2 Majumdar-Ghosh chain, the spin-1 $J_1-J_2$ Heisenberg chain, where $J_2$ is a next-nearest neighbor Heisenberg coupling, and the spin-1 $J_1-J_3$ model, where $J_3$ refers to a three-site interaction defined by $J_3\left[({\bf S}_{i-1}\cdot {\bf S}_i)({\bf S}_i\cdot {\bf S}_{i+1})+\mathrm {H.c.}\right]$. Although both models are rigorously equivalent to the Majumdar-Ghosh chain for spin-1/2, their physics appears to be quite different for spin 1. Indeed, when all couplings are antiferromagnetic, the spin-1 $J_1-J_2$ model undergoes an effective decoupling into two next-nearest neighbour (NNN) Haldane chains upon increasing $J_2$, while the $J_1-J_3$ chain undergoes a spontaneous dimerization similar to the spin-1/2 Majumdar-Ghosh chain upon increasing $J_3$. By extending the phase diagram to all signs of the couplings, we show that both the dimerized and the NNN-Haldane phase are actually present in the $J_1-J_3$ model, the former one adjacent to the Haldane one, the latter one to the ferromagneric one, with an Ising transition between them. By contrast, the $J_1-J_2$ chain only has a NNN-Haldane phase between the Haldane phase and the ferromagnetic phase for positive $J_2$. In both cases, our DMRG data are consistent with a continuous Kosterlitz-Thouless transition between the NNN-Haldane and the ferromagnetic phases.
\end{abstract}

\maketitle


\section{Introduction}

Antiferromagnetic  Heisenberg  spin  chains  have  been studied  intensively  over  the  years. Frustration in one-dimensional systems introduced through competing interactions is known to induce new phases and quantum phase transitions. One of the well known examples is the spin-1/2 chain with nearest and next-nearest neighbor interactions, also known as the Majumdar-Ghosh chain\cite{MajumdarGhosh}, described by the Hamiltonian
\begin{equation}
  H_{J_1-J_2}=\sum_i J_1{\bf S}_i\cdot{\bf S}_{i+1}+J_2{\bf S}_{i-1}\cdot {\bf S}_{i+1}
  \label{eq:j1j2}
\end{equation}
It undergoes a Kosterlitz-Thouless transition between a critical phase and a spontaneously dimerized phase when the ratio between the next-nearest-neighbor and  nearest-neighbor couplings is equal to $J_2/J_1\approx0.2411$\cite{MajumdarGhosh,okamoto}. 

Quite remarkably, there are two different spin-1 generalizations of the Majumdar-Ghosh model. The most obvious one is the spin-1 version of the $J_1-J_2$ model of Eq.\ref{eq:j1j2}. However, for spin-1, it has been shown that there is another model with an interaction term involving next-nearest neighbors that reduces to the $J_1-J_2$ model for spin-1/2, namely the $J_1-J_3$ model defined by the Hamiltonian\cite{michaud1}
\begin{equation}
  H_{J_1-J_3}=\sum_i J_1{\bf S}_i\cdot{\bf S}_{i+1}+J_3\left[({\bf S}_{i-1}\cdot {\bf S}_i)({\bf S}_i\cdot {\bf S}_{i+1})+\mathrm {H.c.}\right].
  \label{eq:j1j3}
\end{equation} 
For spin-1/2, this Hamiltonian reduces to the $J_1-J_2$ model because Pauli matrices anticommute and square to the identity, but for spin-1, this is a different model. 

In both cases, the additional term competes with the nearest-neighbor interaction, but the frustration induced by this competition has quite different consequences. In the spin-1 $J_1-J_2$ model, the next-nearest-neighbor interaction induces a first-order phase transition into a phase that consists of two effectively decoupled next-nearest-neighbors (NNN) Haldane chains\cite{kolezhuk_prl,kolezhuk_connectivity}. This result, first confirmed numerically for positive $J_1$ and $J_2$ couplings by the density matrix renormalization group (DMRG)\cite{dmrg1,dmrg2,dmrg3,dmrg4}, has been recently extended to ferromagnetic $J_1$ coupling, and the NNN-Haldane phase was found to persist for ferromagnetic $J_1$ until the system undergoes  a phase transition at $J_1/J_2=-4$ to a ferromagnetic phase\cite{lee}, a conclusion also confirmed by our extensive DMRG calculations on larger system sizes (see below).

By contrast, the $J_3$ interaction has been shown to induce a  spontaneous dimerization\cite{michaud1}. In that respect, the $J_1-J_3$ model is a more faithful generalization of the spin-1/2 Majumdar-Ghosh model. This transition is in the WZW SU(2)$_2$ universality class, and it is in all respects analogous to the dimerization that has been reported a long time ago in the spin-1 chain with negative biquadratic coupling, in which case the critical point is integrable by Bethe ansatz\cite{takhtajan,babujian} and could be shown to belong to the SU(2)$_2$ Wess-Zumino-Witten (WZW) universality class\cite{affleck86_1,affleck86_2}. 

In fact, linear combinations of $H_{J_1-J_2}$ and $H_{J_1-J_3}$ are also generalizations of the spin-1/2 Majumdar-Ghosh models, and the phase diagram of the model including both $J_2$ and $J_3$ has been worked out recently for positive couplings\cite{komakura,dim_trans,chepiga_j1j2j3long}. It contains three phases, a Haldane phase, a dimerized phase, and a spontaneously dimerized phase, and the nature of the phase transitions has been fully worked out\cite{dim_trans}. In particular, it has been shown that there is a non-magnetic Ising transition between the NNN-Haldane phase and the spontaneously dimerized phase, a result also reported for the model with $J_2$ and biquadratic interactions\cite{chepiga_comment}.

The presence of two possible phases - NNN-Haldane and spontaneously dimerized - between the natural phases of the spin-1 Heisenberg chain - Haldane for antiferromagnetic $J_1$ positive, ferromagnetic for $J_1$ negative - calls for further investigation. So far, general phase diagrams with all signs of couplings have only been worked out for the $J_1-J_2$ model\cite{lee} and the bilinear-biquadratic model\cite{uimin,lai,sutherland,takhtajan,babujian,haldane1983,haldane1983PRL,PRL_AKLT,barber1989,kluemper1989,Xian1993,fath1991,fath1993,Schollwock1996,buchta2005,PRB_Lauchli}, with the conclusion that there is only a NNN-Haldane phase for the $J_1-J_2$ model, and, for negative biquadratic coupling, only a dimerized phase with a direct transition into the ferromagnetic phase. This transition between the dimerized phase and the ferromagnetic phase is however very special because it takes place at an SU(3) integrable point at which magnetic and quadrupolar states are degenerate. In the case of the $J_1-J_3$ model, we do not expect this to be the case. So we have decided to study its full phase diagram to investigate the fate of its dimerized phase upon approaching the ferromagnetic one. As we shall see, the transition is in that case a two-step process, with first an Ising transition into a NNN-Haldane phase, and then a transition from that NNN-Haldane phase into the ferromagnetic phase. The generic phase diagram of the spin-1 chain with interactions extending to next-nearest neighbors thus appears to contain both a NNN-Haldane phase and a dimerized phase. This very rich phase diagram is fully embodied by the $J_1-J_3$ model, but not by the $J_1-J_2$ model. 

The rest of the paper is organized as follows. Sections \ref{sec:PD} to \ref{sec:KT} are devoted to the $J_1-J_3$ model, while section \ref{sec:j1j2} discusses the $J_1-J_2$ model. We start with a brief overview of the phase diagram in Sec.\ref{sec:PD} and analyze the ground-state energy for the entire parameter range. In Sec.\ref{sec:NNN_Ising} we discuss the appearance of a NNN-Haldane phase and the Ising transition that separates it from the dimerized phase. In Sec.\ref{sec:KT} we investigate the incommensurate short-range order in the NNN-Haldane phase and the phase transition to the ferromagnetic phase. In particular, we compare our DMRG results with the analytical calculation based on a single-magnon instability from the ferromagnetic phase. In sec.\ref{sec:j1j2} we discuss the full phase diagram of spin-1 $J_1-J_2$ model. Finally, in Sec.\ref{sec:conc} we summarize our results and discuss directions for future work.

\section{Phase diagram of the Heisenberg chain with three-site interaction}
\label{sec:PD}
\subsection{Overview}

We start with a brief overview of the general properties of the phase diagram of the $J_1-J_3$ model defined by the Hamiltonian of Eq. \ref{eq:j1j3}.
The following parametrization is introduced for convenience: $J_1=J\cos\theta$ and $J_3=J\sin\theta$ with $\theta\in[0, 2\pi]$. Without loss of generality we set $J=1$.

\begin{figure}[t!]
\includegraphics[width=0.45\textwidth]{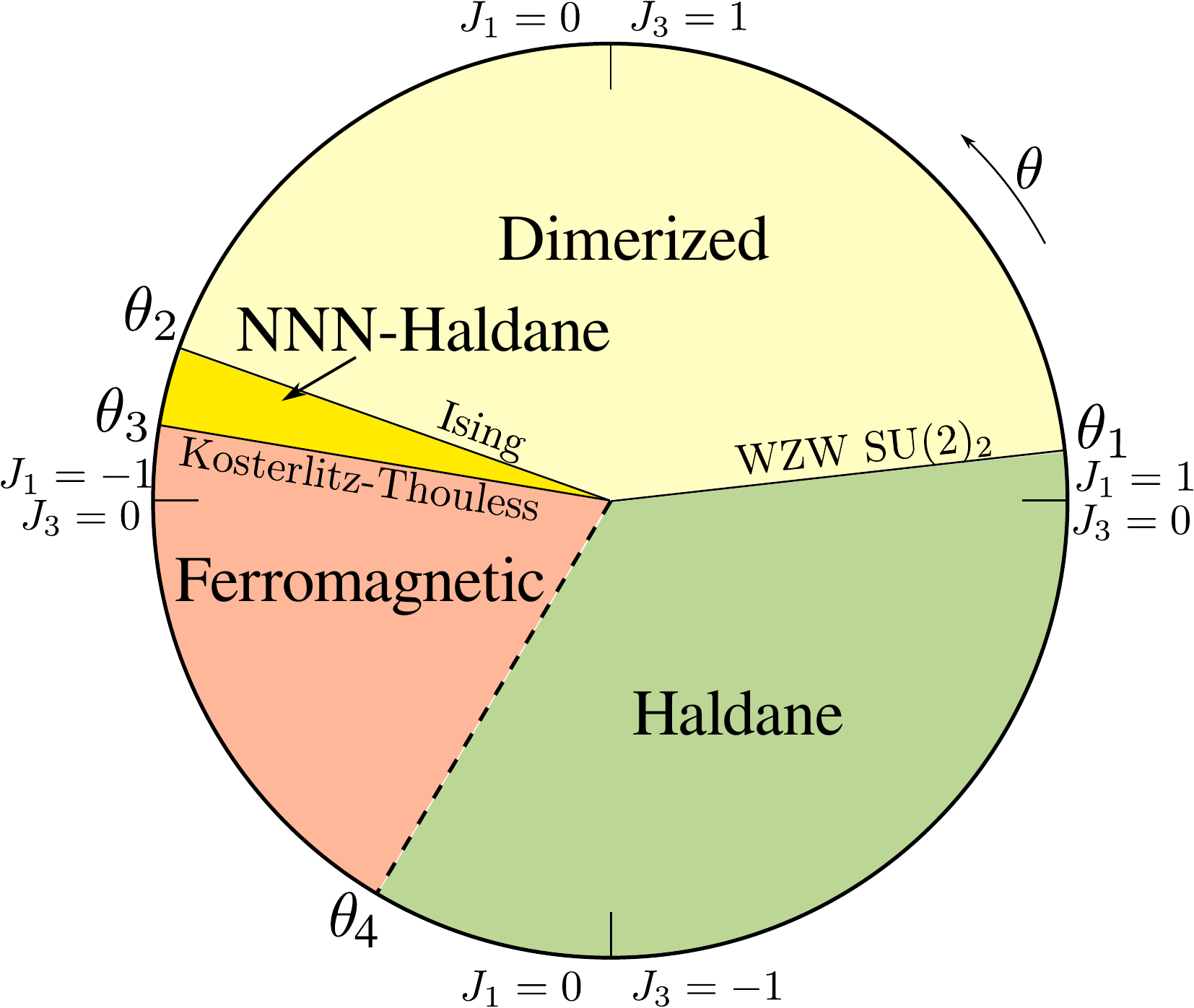}
\caption{Phase diagram of the $J_1-J_3$ model defined by the Hamiltonian of Eq.\ref{eq:j1j3}. The transition between the Haldane and dimerized phases at $\theta_1\approx 0.03519\pi$ is continuous in the WZW SU(2)$_2$ universality class\cite{michaud1}. The transition between the dimerized and NNN-Haldane phases is Ising and takes place at $\theta_2\approx 0.8913\pi$. Kosterlitz-Thouless transition to the ferromagnetic phase takes place at $\theta_3\approx0.9474\pi$. The transition between the ferromagnetic and Haldane phases at $\theta_4\approx1.33\pi$ is first order.}
\label{fig:j1j3_pd}
\end{figure}

The phase diagram as a function of $\theta$ is shown in Fig.\ref{fig:j1j3_pd} and consists of four phases: Haldane, dimerized, NNN-Haldane and ferromagnetic.

The portion of the phase diagram with positive couplings $J_1$ and $J_3$ has been first studied in Ref.\onlinecite{michaud1}, with the following conclusions:
{\it i)} for small values of the parameter $\theta$ the system is in the Haldane phase. Upon increasing $\theta$, the three-site interaction induces spontaneous dimerization \cite{michaud1}; {\it ii)} the transition between the two phases is located at  $J_3/J_1\simeq0.111$ ($\theta\approx 0.0352\pi$) \cite{michaud1}. It is continuous, and the critical point is in the WZW SU(2)$_2$ universality class;  {\it iii)} there is an exactly dimerized point at $J_3/J_1=1/6$ ($\theta\approx 0.0526\pi$) that can be viewed as a generalization of the Majumdar-Ghosh point \cite{MajumdarGhosh} to larger spins\cite{michaud1}. This point coincides with the disorder point where the spin-spin correlations become incommensurate \cite{dim_trans}.

According to our numerical data, the dimerized phase remains stable also for ferromagnetic values of the nearest-neighbor coupling.
Inside the dimerized phase, there is a transition between the two possible ground states  with open boundary conditions: for $\theta<0.55\pi$ a strong dimer is formed on the first bond, while for $\theta\geq 0.55\pi$ a strong dimer is formed on the second bond. So, in finite chain with open boundary conditions the orientation of the dimers changes at this point. As a consequence, spin-1 edge states appear for $\theta\geq 0.55\pi$.

At $\theta_2\approx0.8913\pi$ the system undergoes a quantum phase transition from the dimerized to the NNN-Haldane phase. The transition is continuous in the Ising universality class in agreement with previous results for antiferromagnetic spin-1 chains\cite{dim_trans,chepiga_comment}. In the thermodynamic limit the singlet-triplet bulk gap remains open, while the spectrum becomes critical within the singlet sector.
 Due to the ferromagnetic nature of the nearest-neighbor coupling, the system has emergent spin-1 edge states in analogy with the edge states detected previously for the $J_1-J_2$ chain when the $J_1$ coupling is ferromagnetic\cite{zero_modes_long}.

Beyond $\theta_3=\arctan(-1/6)$ the system is in the ferromagnetic phase. Our results are consistent with a direct Kosterlitz-Thouless phase transition between the NNN-Haldane and the ferromagnetic phases.
The transition between the ferromagnetic phase and the Haldane phase is first order and it takes place at $\theta_4\approx 1.33\pi$.

In the next sections we explain in details how the phase diagram of Fig.\ref{fig:j1j3_pd} has been obtained.

\subsection{Energy}

\begin{figure}[h!]
\centering 
\includegraphics[width=0.4\textwidth]{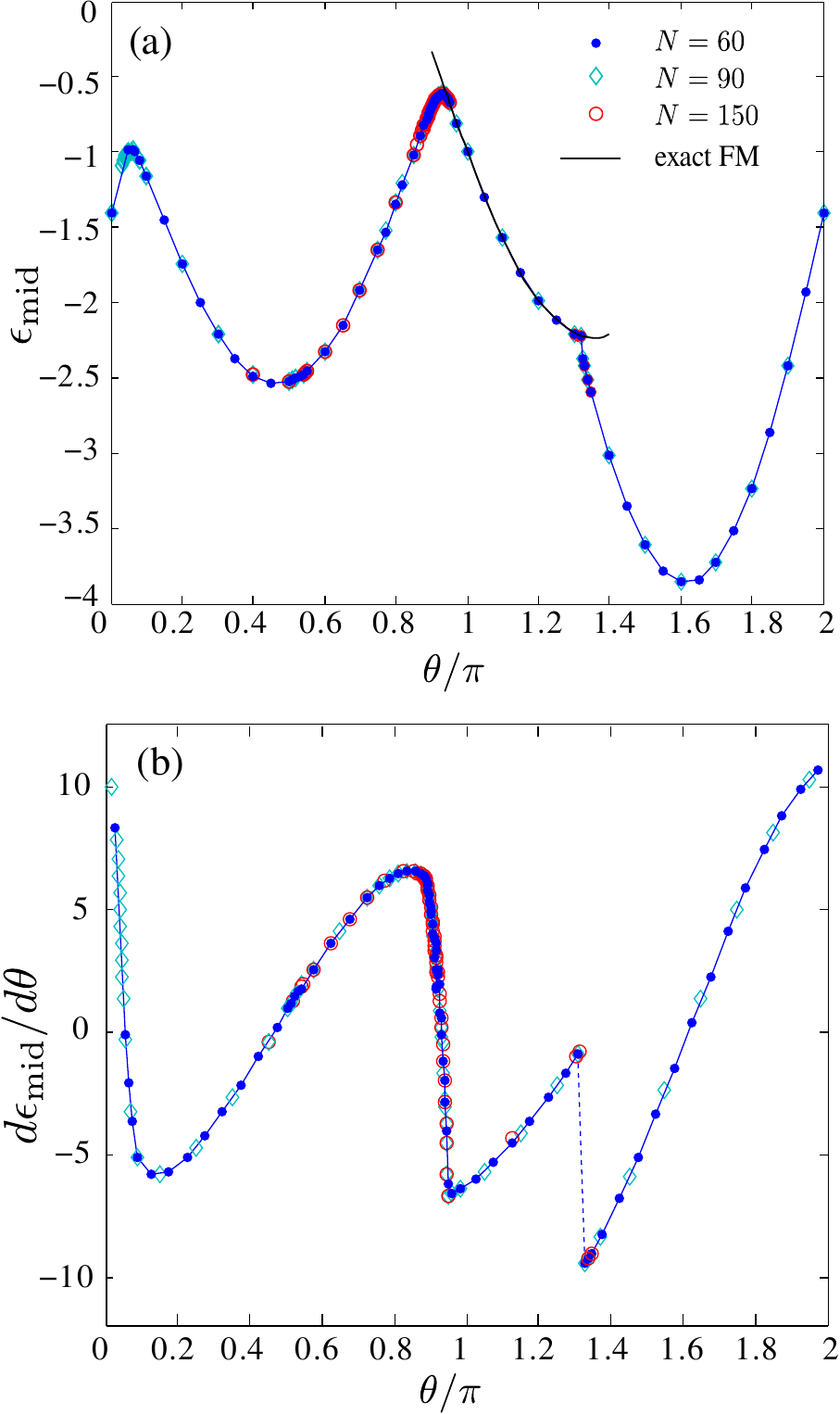}
\caption{Ground-state energy per bond calculated in the middle of finite-size chains as a function of $\theta$. In the ferromagnetic phase, the DMRG data coincide exactly with the analytic prediction $\epsilon_{FM}=\cos\theta+2\sin\theta$. }
\label{fig:j1j3_energy}
\end{figure}

The ground-state energy per bond calculated in the middle of the chain is shown in Fig.\ref{fig:j1j3_energy}(a).
The ferromagnetic phase can be well described by the classical state with all spins aligned. The classical energy per site of this state is given by $\epsilon_{FM}=J_1+2J_3=\cos\theta+2\sin\theta$.
 The pronounced kink around $\theta\approx 1.33\pi$ signals a first order phase transition between the ferromagnetic phase and the Haldane phase as indicated in the phase diagram of Fig.\ref{fig:j1j3_pd}. 

We have also calculated the first derivative of the energy per site with respect to $\theta$. The results are shown in Fig.\ref{fig:j1j3_energy}(b).
In such a plot, a first order phase transition is clearly seen as a finite jump in the derivative. The kink in the derivative of the energy around $\theta_3\approx 0.947\pi$ suggests a second order phase transition to the ferromagnetic phase. One can also see a slight kink in the derivative around $\theta=0.9\pi$. As we shall see, it corresponds to an Ising transition between the dimerized and NNN-Haldane phases.

\section{NNN-Haldane phase and Ising transition}
\label{sec:NNN_Ising}

\begin{figure}[h!]
\centering 
\includegraphics[width=0.49\textwidth]{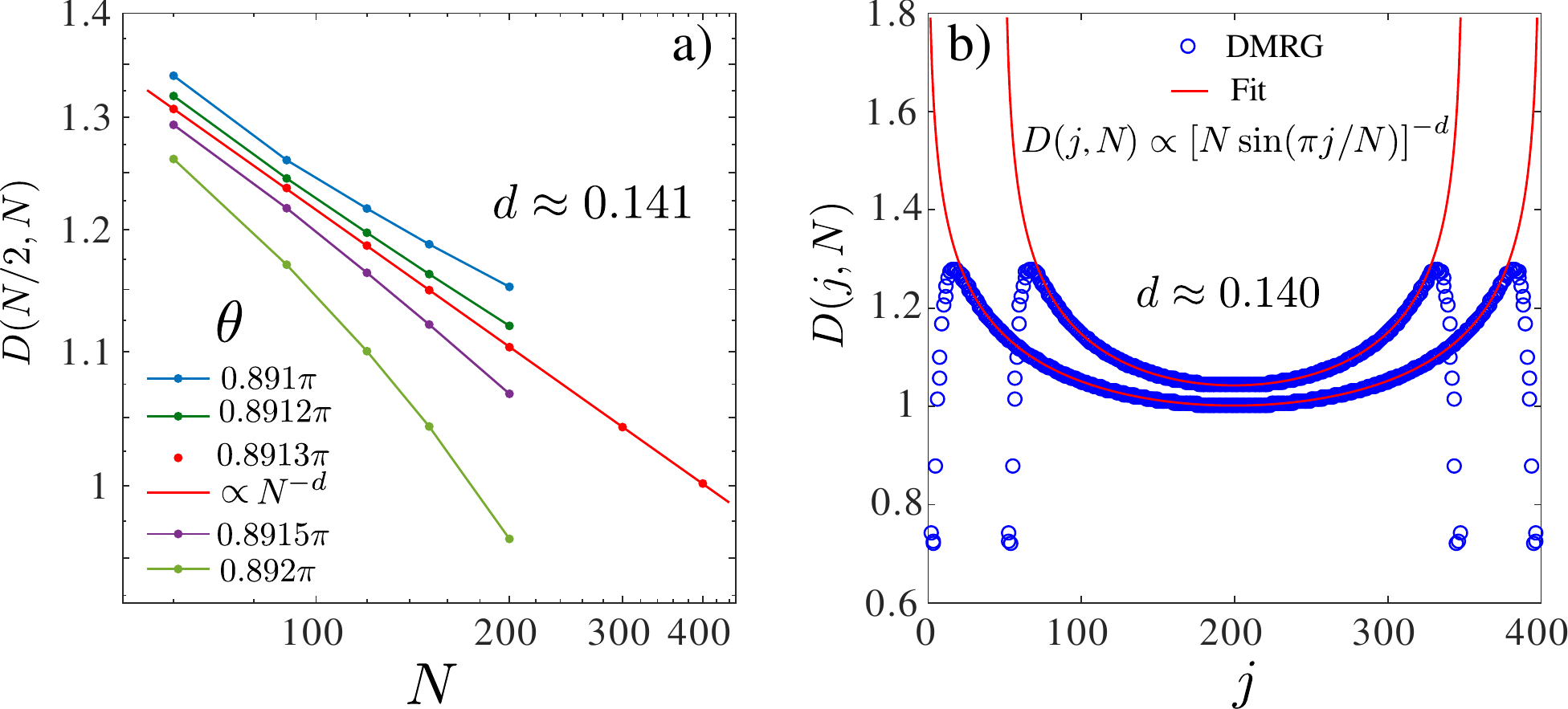}
\caption{Scaling of the dimerization at the critical Ising point. (a) Log-log plot of the dimerization. The linear curve corresponds to the Ising critical point, and the slope to the critical exponent. This leads to $\theta_c=0.8913\pi$, and to a slope $0.141$, that agrees within $15\%$ with the prediction $1/8$ for Ising critical point. (b) Site dependence of $D(j,N)$ at the critical point fitted to $1/[N \sin (\pi j/N)]^{d}$.
This leads to an exponent $d=0.140$, that also agrees with the Ising prediction $1/8$ within $15\%$. }
\label{fig:j1j3_dim}
\end{figure}

In order to distinguish the dimerized phase from the non-dimerized one, we introduce the dimerization parameter $D(j,N)=|\langle \vec S_j\cdot\vec S_{j+1} \rangle - \langle \vec S_{j-1}\cdot\vec S_j \rangle|$. We plot the values of the middle-chain dimerization ($j=N/2$) as a function of the system size in a log-log scale. The phase transition is then associated with the separatrix. The slope gives the critical exponent. In Fig.\ref{fig:j1j3_dim}(a) we present our numerical results for the dimerization around the transition located at $\theta\approx0.8913\pi$. The critical exponent extracted from the scaling of the middle chain dimerization is $d\approx0.141$. It agrees within $15\%$ with the corresponding Ising critical exponent $1/8$.

This critical exponent can also be extracted from the scaling of the dimerization along a finite-size chain. According to conformal field theory, at the  critical point it scales as $D(j,N)\propto 1/[N \sin (\pi j/N)]^{d}$. The fit of our numerical data for $N=300$ and $N=400$ is shown in Fig.\ref{fig:j1j3_dim}(b). Since edge effects are very strong, we exclude from the fit at least 25 points at each edge. The obtained critical exponent $d\approx0.140$ agrees with the one computed above.

\begin{figure}[h!]
\centering 
\includegraphics[width=0.45\textwidth]{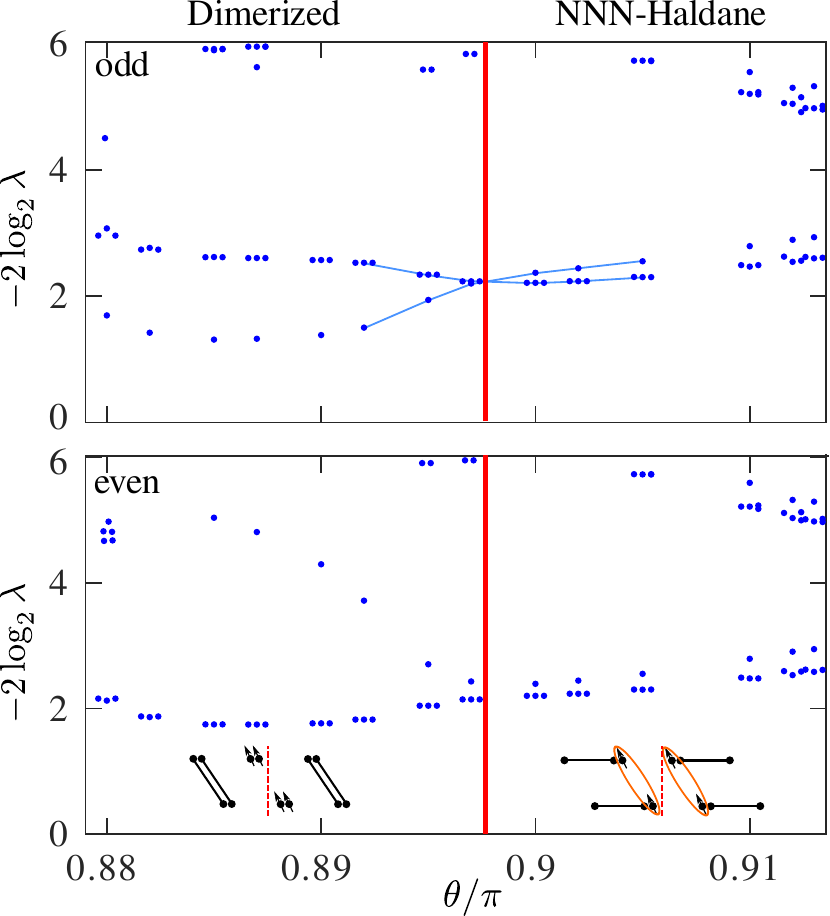}
\caption{Entanglement spectrum for an open chain with $N=150$ sites as a function of $\theta$ when cut across (a) an odd and (b) an even bond. Only the lower part of the spectrum is shown. The dots show the multiplicity of the Schmidt values. Insets: VBS sketches of the various boundaries created by the cut of the chain inside the dimerized and NNN-Haldane phases. The two spins $1/2$ created at each edge in the NNN-Haldane phase couple with each other and form a triplet, which is represented as an orange ellipse. This leads to the three-fold degeneracy of the lowest state of the entanglement spectrum.}
\label{fig:j1j3_entspec}
\end{figure}

 In order to understand the nature of the non-dimerized phase, we computed the entanglement spectrum as a function of $\theta$. The entanglement spectra for a system cut across odd and even bonds are shown in Fig.\ref{fig:j1j3_entspec} (a) and (b) respectively.  As explained above, for $\theta>0.55\pi$ the edges favor a bond without dimer
and therefore every even bond is occupied by a singlet dimer built our of spins 1.  Thus the lowest level in the entanglement spectra is threefold degenerate when the system is cut across an even bond and non-degenerate otherwise.

In the non-dimerized phase the lowest level in the entanglement spectrum is always three-fold degenerate. Therefore the ground state is translationally invariant by one site. We interpret this non-dimerized phase as two next-nearest-neighbor Haldane chains ferromagnetically coupled by the $J_1$ interaction.
In the case of antiferromagnetic nearest-neighbor coupling, the entanglement spectrum in the NNN-Haldane phase is non-degenerate since the two spins $1/2$ created at each edge when the system is cut couple with each other and form a singlet. By contrast, the ferromagnetic coupling between the two spins $1/2$ at the edge induces a triplet, and the degeneracy of the entanglement spectrum is three.

\begin{figure}[h!]
\centering 
\includegraphics[width=0.48\textwidth]{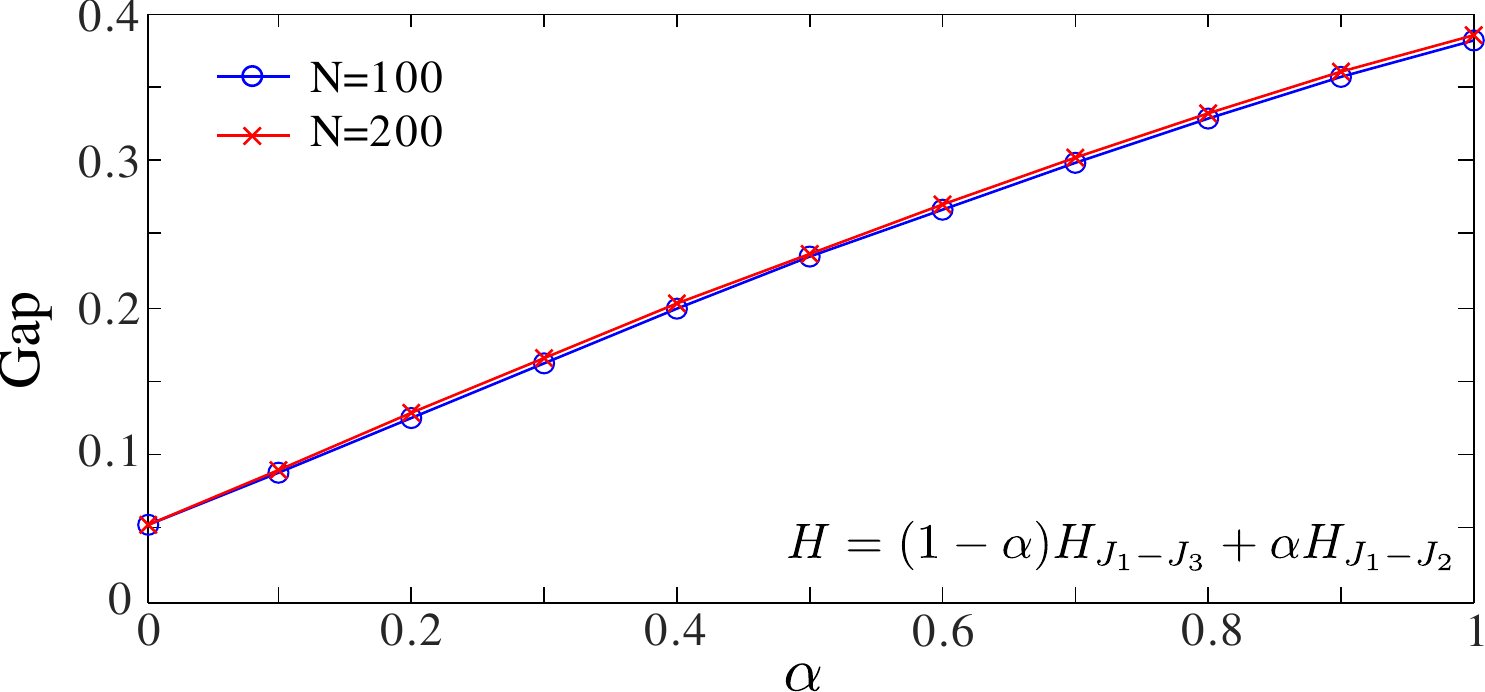}
\caption{Energy gap while interpolating between the $J_1-J_3$ model ($\alpha=0$) with $J_3/J_1=\tan(0.9\pi)$ and the $J_1-J_2$ model ($\alpha=1$) with $J_1=-J_2$, which is known to be in the NNN-Haldane phase. Since the gap remains open for all the values of the interpolation parameter $\alpha$, the non-dimerized phase in the $J_1-J_3$ model is identified as a NNN-Haldane phase.}
\label{fig:j1j3_interpolation}
\end{figure}

In order to confirm our interpretation of the non-dimerized phase, we interpolate between the $J_1-J_3$ model and the model with ferromagnetic nearest-neighbor ($J_1<0$) and antiferromagnetic next-nearest-neighbor ($J_2>0$)  interactions, for which the appearance of the NNN-Haldane phase is natural and has been confirmed recently in Ref.\onlinecite{lee}. We consider the following Hamiltonian:
\begin{multline}
  H_{J_1J_3\rightarrow J_1J_2}=\cos\theta\sum_i [{\bf S}_i\cdot{\bf S}_{i+1}-\alpha{\bf S}_i\cdot{\bf S}_{i+2}]+\\
(1-\alpha)\sin\theta\sum_i\left[({\bf S}_{i-1}\cdot {\bf S}_i)({\bf S}_i\cdot {\bf S}_{i+1})+\mathrm {H.c.}\right],
  \label{eq:j1j2interp}
\end{multline}
where $\alpha$ interpolates between the $J_1-J_3$ and the $J_1-J_2$ models. When $\alpha=0$, the Hamiltonian of Eq.\ref{eq:j1j2interp} reduces to the $J_1-J_3$ model with our usual parametrization $J_1=\cos\theta$ and $J_3=\sin\theta$. Here we take $\theta=0.9\pi$,  inside the NNN-Haldane phase. When $\alpha=1$, the last term vanishes and the Hamiltonian reduces to the $J_1-J_2$ model with $J_1=-J_2=\cos\theta$. This point is located inside the NNN-Haldane phase (see Fig.\ref{fig:phase_diag_j1j2} discussed below). According to Fig.\ref{fig:j1j3_interpolation}, the gap, calculated within the symmetry sector of the ground state, never closes while interpolating between the $J_1-J_3$ and $J_1-J_2$ models with the parameters specified above. Therefore, the non-dimerized gapped phase that appears at $\theta=0.8913\pi$ for the  $J_1-J_3$ model can indeed be identified as a NNN-Haldane phase.

Let us discuss in more details the nature of the phase transition between this NNN Haldane phase and the dimerized phase. In the case of antiferromagnetic nearest neighbor coupling, the transition between the NNN-Haldane phase and the dimerized phase in the $J_1-J_2-J_3$ model is in the Ising universality class\cite{dim_trans,chepiga_comment}. This transition occurs entirely in the singlet sector while the singlet-triplet gap remains open, a consequence of the fact that the domain walls between these phases  are non-magnetic\cite{dim_trans}. 
The critical exponent $d\approx0.14$ extracted from the scaling of the dimerization parameters for the $J_1-J_3$ model as shown in Fig.\ref{fig:j1j3_dim}
 only agrees within $13\%$ with the Ising prediction $d=1/8$.
 The relatively big discrepancy between the critical exponent computed numerically and the field theory prediction presumably comes from strong edge effects. Indeed, as shown in Fig.\ref{fig:j1j3_dim}(b), the dimerization decreases close to the edges. In Ising boundary conformal field theory there are only two types of conformally-invariant boundary conditions - either free or fixed, while partially polarized boundary condition re-normalizes to fixed boundary conditions with some corrections. We believe that these corrections are the main source of discrepancy between the numerical value of the critical exponent and the conformal field theory prediction.  

\begin{figure}[h!]
\centering 
\includegraphics[width=0.47\textwidth]{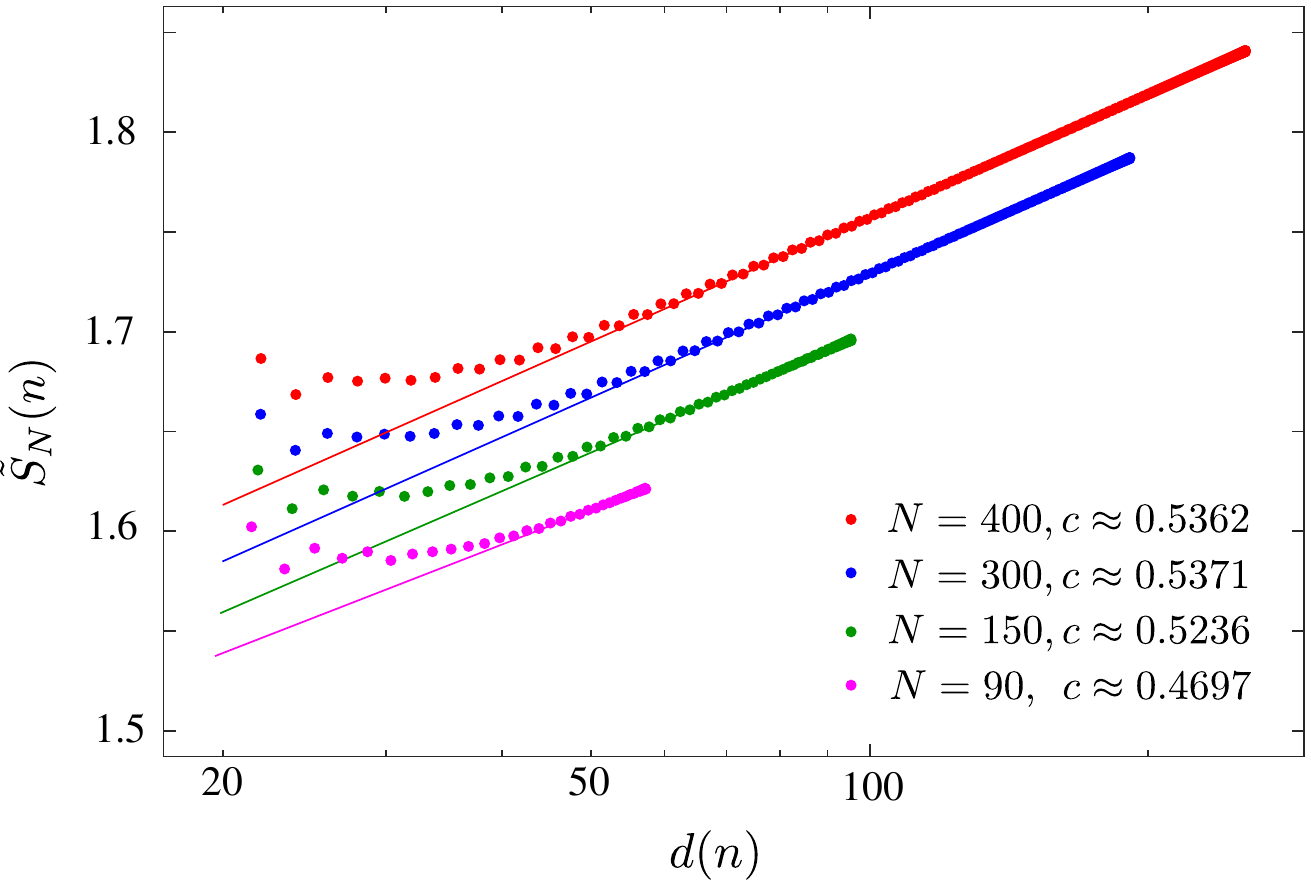}
\caption{Scaling of the entanglement entropy of open chains after removing the Friedel oscillations with conformal distance $d(n)$ for $\theta=0.8913\pi$ and different system sizes. The extracted values of the central charge agree within $8\%$ with the field-theory prediction $c=1/2$ for the Ising critical theory. Data are shifted vertically for clarity.}
\label{fig:j1j3_central charge}
\end{figure}

It is well established that the DMRG algorithm has better performances for open systems, and that much bigger system sizes can be reached than for periodic boundary conditions. In systems with open boundary conditions, the entanglement entropy scales with the block size according to:
\begin{equation}
S_N(n)=\frac{c}{6}\ln\left[\frac{2N}{\pi}\sin\left(\frac{\pi n}{N}\right)\right]+s_1+\ln g
\label{eq:calabrese_cardy_obc}
\end{equation}
Since we are dealing here with much larger system sizes, it is useful to present the results in logarithmic scale by introducing the conformal distance $d$:
\begin{equation}
d=\frac{2N}{\pi}\sin\left(\frac{\pi n}{N}\right)
\label{eq:conformal_distance}
\end{equation}
On top of the prediction of Eq.~\ref{eq:calabrese_cardy_obc} big oscillations appear due to Friedel oscillations present in systems with open boundary conditions. 
In order to remove these oscillations, following Ref.\onlinecite{capponi}, we have subtracted the spin-spin correlation on the corresponding link from the entanglement entropy with some weight $\zeta$. Then the reduced entanglement entropy as a function of the conformal distance takes the form:
\begin{equation}
\tilde{S}_N(n)=\frac{c}{6}\ln d(n)+\zeta \langle {\bf S}_n{\bf S}_{n+1} \rangle+s_1+\ln g
\label{eq:calabrese_cardy_obc_corrected}
\end{equation}
An example of fit of the entanglement at the critical point with the Calabrese-Cardy formula of Eq. \ref{eq:calabrese_cardy_obc_corrected} is provided in Fig.\ref{fig:j1j3_central charge}. The resulting values of the central charge agree within $8\%$ with the field theory prediction $c=1/2$ for Ising.

As a further confirmation of the Ising universality class of the critical point between the dimerized and the NNN-Haldane phases, we have computed the finite-size scaling of the ground-state energy and excitation energies for open chains with even and odd numbers of sites. Since both open edges favor  bonds without dimers, we associate the spin-1 chain with an even (odd) number of sites with Ising $\uparrow,\uparrow$ ($\uparrow,\downarrow$) boundary condition, in complete analogy with the case of antiferromagnetic nearest-neighbor coupling \cite{dim_trans,chepiga_j1j2j3long}.

\begin{figure}[h!]
\centering 
\includegraphics[width=0.49\textwidth]{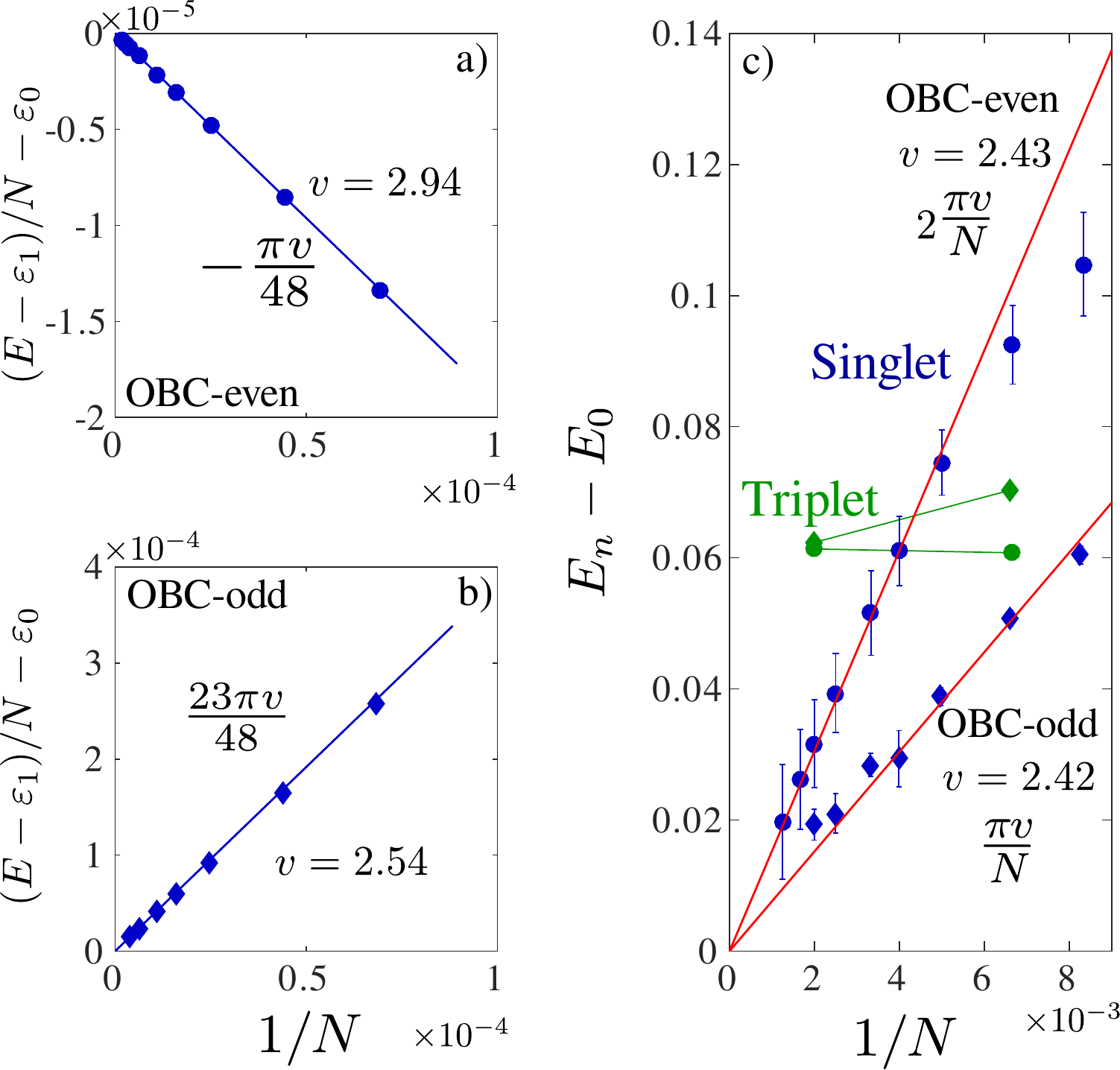}
\caption{Left panels: Linear scaling of the ground state energy per site in open chains with $1/N^2$ after subtracting $\varepsilon_0$ and $\varepsilon_1$ terms for (a) even and (b) odd number of sites. The extracted values of the velocities agree with each other within $15\%$. Right panel: Energy gaps in the singlet (blue) and triplet (green) sectors for open boundary conditions as a function of $1/N$ for even (circles) and odd (diamonds) number of sites }
\label{fig:j1j3_conftow}
\end{figure}

The ground-state energy of the system with conformally invariant boundary conditions at the Ising critical point scales according to:
\begin{equation}
  E=\varepsilon_0N+\varepsilon_1+{\pi v\over N}\left[-{1\over 48}+x\right],
\end{equation}
where $\varepsilon_0$ and $\varepsilon_1$ are non-universal coupling constants and $x$ is the corresponding conformal dimension. As predicted by Cardy\cite{Cardy89},  the energy spectrum is given by the identity conformal tower $I$ for $\uparrow ,\uparrow$ boundary conditions, and by the $\epsilon$ conformal tower for $\uparrow ,\downarrow$ boundary conditions. The corresponding scaling dimensions are $x_{I}=0$ and $x_{\epsilon}=1/2$. The finite-size scaling of the numerically obtained values of the ground-state energy for even and odd number of sites are shown in Fig.\ref{fig:j1j3_conftow}(a) and (b). The extracted values of the velocities agree with each other within $15\%$.

We have also calculated the excitation energy of the lowest excited state in the sector $S^z_\mathrm{tot}=0$ (see Fig. \ref{fig:j1j3_conftow}(c)). In order to extract excitation energies with DMRG at the critical point, we have followed the procedure described in Ref.\onlinecite{dmrg_chepiga}.
 The bulk gap scales linearly with $1/N$, in agreement with a conformally invariant critical theory. Moreover, according to conformal field theory, the lowest excitation energy that belongs to the $\epsilon$ conformal tower scales with the system size as $\pi v/N$, while for the $I$ conformal tower, the lowest excitation energy scales as $2\pi v/N$.

The low-lying excitations shown in green in Fig.\ref{fig:j1j3_conftow}(c) have been computed in the sector $S^z_\mathrm{tot}=0$ and are localized edge excitations. 
Due to the spin-1 edge states induced by the ferromagnetic $J_1$ coupling, three low-lying states are expected - singlet, triplet and quintuplet. Depending on the size of the system, one of them becomes the ground state, and the other two are separated by an exponentially small gap.
The singlet-triplet gap can be thus obtained as the difference between the ground-state energy and the third state in the sector of $S^z_\mathrm{tot}=0$. 
However, one can neglect the exponentially small gap for long enough chains.
Then effectively the singlet-triplet gap  can be obtained as the lowest state in the sector $S^z_\mathrm{tot}=3$ since this state can be thought of as a bulk triplet combined to a spin-2 in-gap state of negligible energy. This is much simpler to detect from the numerical point of view. The gap between the lowest states in the sectors $S^z_\mathrm{tot}=3$ and $S^z_\mathrm{tot}=0$ is in good agreement with the localized edge excitations obtained by calculating several energy levels in the sector $S^z_\mathrm{tot}=0$. Therefore we interpret these excitations as triplet excitations. Since these localized edge excitations occur at both edges, the corresponding energy levels are two-fold degenerate.
Note also that this singlet-triplet gap is small, and the first singlet excitation appears below the triplet one at relatively large system sizes ($N>120$ resp. $N>250$ for chains with odd resp. even total number of sites).

As pointed out above, one of the arguments in favor of an Ising transition in the context of antiferromagnetic $J_1-J_2-J_3$ chain is the non-magnetic character of the domain wall between the NNN-Haldane and the dimerized phases\cite{dim_trans}. In the present case, due to the ferromagnetic nearest-neighbor coupling, a pair of spins-1/2 at the edge of a domain in the NNN-Haldane phase does not form a singlet, but a triplet, as can be deduced from the entanglement spectra  of Fig.\ref{fig:j1j3_entspec}. However, a positive $J_1$ also favors a state without a dimer at the edge of the dimerized phase. 

\begin{figure}[h!]
\centering 
\includegraphics[width=0.47\textwidth]{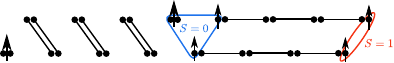}
\caption{Sketch of the non-magnetic domain wall between the NNN-Haldane phase and the dimerized phase.}
\label{fig:j1j3_domainwall}
\end{figure}

The fact that the transition is in the Ising universality class leads us to speculate that the spins-1 that appear at the boundaries of the Haldane and the dimerized domains couple to each other and form a singlet, leading to a non-magnetic domain wall between these two phases, as sketched in  Fig.\ref{fig:j1j3_domainwall}.

\begin{figure}[h!]
\centering 
\includegraphics[width=0.47\textwidth]{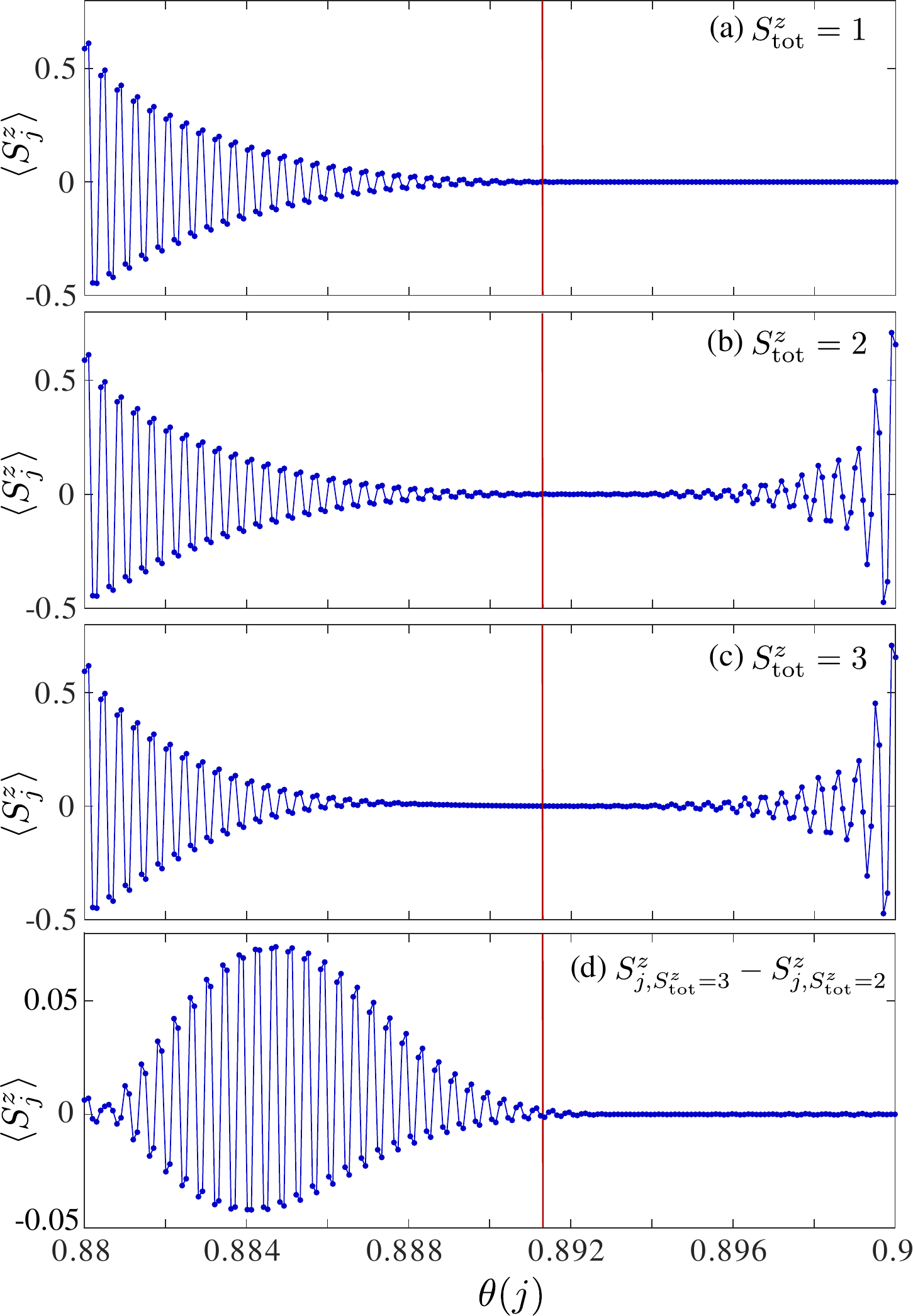}
\caption{Local magnetization along a chain with $N=200$ sites and as a function of the parameter $\theta$ that changes linearly along the chain interpolating between the dimerized phase on the left and NNN-Haldane phase on the right. (a) A spin 1 is localized at the dimerized edge of the chain. (b) A spin 1 is localized at each edge. (c) A spin 1 is localized at each edge and, as follows from (d), there is also a spin 1 idelocalized over the dimerized domain. (d) shows the difference between (c) and (b).}
\label{fig:j1j3_gradient}
\end{figure}

In order to verify the non-magnetic nature of the domain wall, we have used gradient DMRG - a finite-size version of the DMRG where the coupling constants vary linearly along the chain. This approach, also known as DMRG scans, is traditionally used for 2D DMRG to probe various phases and approximately locate the critical lines\cite{PhysRevLett.120.207203,PhysRevB.98.085104}. Here we propose to use it to probe the domain wall between two corresponding phases. We consider a chain with $N=200$ sites and we linearly increase the parameter $\theta$ of the local Hamiltonian (matrix product operator) from the dimerized phase at $0.88\pi$ to the NNN-Haldane phase at $0.9\pi$. We set the total magnetization of the chain to be $S^z_\mathrm{tot}=1,2,3$, and we measure the local magnetization along the chain. According to Fig.\ref{fig:j1j3_gradient}, for $S^z_\mathrm{tot}=1$ a spin-1 is localized at the left edge. It corresponds to the edge state of the dimerized phase. When $S^z_\mathrm{tot}=2$, spin-1 edge states are localized at both ends of the chain. The energy difference between the states with $S^z_\mathrm{tot}=1$ and $S^z_\mathrm{tot}=2$ and the ground-state with $S^z_\mathrm{tot}=0$ is less than $10^{-5}$.
   For $S^z_\mathrm{tot}=3$,  in addition to two spins-1 localized at each edge, another spin-1 is delocalized over the domain of the dimerized phase. It partially compensates the magnetization induced by the spin-1 at the left edge. In order to see the nature of this additional spin-1 more clearly we plot the difference between the local magnetizations in the sector $S^z_\mathrm{tot}=3$ and $S^z_\mathrm{tot}=2$. The resulting magnetization is shown in Fig.\ref{fig:j1j3_gradient}(d). It is consistent with a bulk excitation delocalized over the left (dimerized) part of the chain. The gap between the ground states in the sectors $S^z_\mathrm{tot}=3$ and $S^z_\mathrm{tot}=2$ is equal to $\Delta\approx0.039$, orders of magnitudes larger than the energy difference between the in-gap states with lower magnetization. Therefore we one can interpret this state as a triplet bulk excitation. Note, though, that "the bulk" in this case rather means the dimerized domain, while the state of the right part of the chain (NNN-Haldane domain) remains almost unchanged. These results clearly show that there is no magnetization at the boundary between the two domains, hence that the domain wall is non-magnetic.

\section{Transition between the NNN-Haldane and the ferromagnetic phases}
\label{sec:KT}

\subsection{One magnon instability from the ferromagnetic phase}

First of all, we look at the one-magnon instability from the ferromagnetic state. We start with the ferromagnetic wave-function polarized in the $z$-direction
$|0\rangle=|S,S,...,S,S\rangle$, with $S=1$. For any site, $l$ $S_l^+|0\rangle=0$, and . The energy of this state can be obtained by applying the Hamiltonian of Eq. \ref{eq:j1j3} to this state:
\begin{equation}
  E_{FM}=(J_1+2J_3)N=(\cos\theta+2\sin\theta)N
\end{equation}

We consider a single magnon state from the ferromagnetic state $|0\rangle$ defined by:
\begin{equation}
|\varphi(q)\rangle=\frac{1}{\sqrt{2NS}}\sum_{l=1}^Ne^{iql}S^-_l|0\rangle
\label{eq:one_mag}
\end{equation}
where $q$ is the momentum, and $1/\sqrt{2NS}$ is a normalization factor. The energy of this state (see derivation in the Appendix \ref{ap:one_magnon_j1j3}) as a function of $\theta$ and $q$ is given by:

\begin{equation}
E(q)=f_1+2f_2\cos q+2f_3\cos 2q,
\label{eq:j1j2:disp}
\end{equation}
with
\begin{multline}
f_1=(N-2)\cos\theta+2(N+1)\sin\theta\\
f_2=\cos\theta+2\sin\theta\\
f_3=\sin\theta\\
\end{multline}

At $\theta=\pi$ ($J_1=-1$ and $J_3=0$) the system is in the ferromagnetic state, the energy given by Eq.\ref{eq:j1j2:disp} has a minimum at $q=0$ and is equal to the ferromagnetic energy. Upon decreasing $\theta$, the curvature of the energy as a function of $q$ at $q=0$ decreases. The phase transition occurs when the curvature changes sign:
\begin{equation}
C=\left.\frac{dE(q)}{dq}\right|_{q=0}
=-\left.2f_2\cos q -8 f_3\cos q\right|_{q=0}=0
\end{equation}
The solution is
\begin{equation}
\frac{f_2}{f_3}=\frac{J_1+2J_3}{J_3}=-4.
\end{equation}
This leads to $J_3/J_1=-1/6$, i.e. $\theta\approx0.9474\pi$. 
The wave-vector for which the energy is minimized is given by:
\begin{equation}
-2f_2\sin q-4f_3\sin 2q=0.
\end{equation}
The non-trivial solution of this equation is $\cos q=-f_2/4f_3$, leading to:
\begin{equation}
q=\arccos\left(-\frac{1}{2}-\frac{1}{4}\cot\theta \right).
\label{eq:vave_vec_j1j3}
\end{equation}

\subsection{DMRG results in the NNN-Haldane phase}

\begin{figure}[h!]
\centering 
\includegraphics[width=0.47\textwidth]{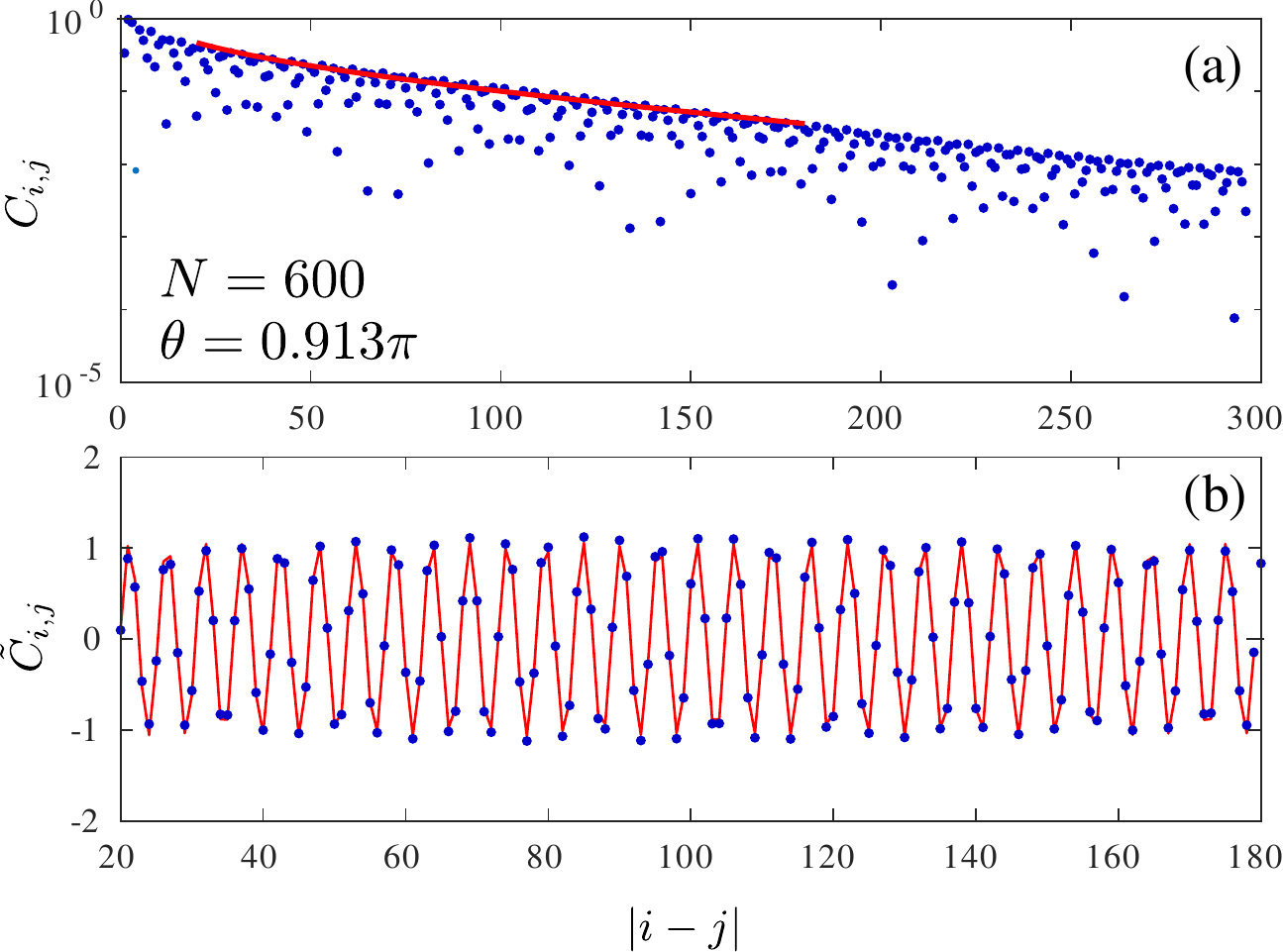}
\caption{ Example of fit of the spin-spin correlations to the Ornstein-Zernicke form given by Eq.\ref{eq:OZ} for $N=600$ and $\theta=0.913\pi$. In the first step (a), we extract the correlation length discarding the oscillations. In the second step (b), we fit the reduced correlation function to extract the wave-vector $q$.}
\label{fig:OZfit_example}
\end{figure}

In order to understand what happens to the system while approaching the ferromagnetic phase we look at the correlation length $\xi$ and the wave-vector $q$ of the incommensurate correlations. 
We extract $\xi$ and $q$ by fitting the spin-spin correlations 
$C_{i,j}=\langle {\bf S}_i \cdot {\bf S}_j \rangle-\langle {\bf S}_i\rangle\cdot\langle{\bf S}_j \rangle$
 to the Ornstein-Zernicke form:

\begin{equation}
  C^\mathrm{OZ}_{i,j}\propto \frac{e^{-|i-j|/\xi}}{\sqrt{|i-j|}}\cos(q|i-j|+\varphi_0),\label{eq:OZ}
\end{equation}
where the correlation length $\xi$, the wave-vector $q$ and the phase shift $\varphi_0$ are fitting parameters. We find that the quality of the fit is improved if it is done in two steps. First, we discard the oscillations and fit the main slope of the exponential decay as shown in Fig.\ref{fig:OZfit_example}(a). This allows us to perform a fit in a semi-log scale $\log C(x=|i-j|)\approx c-x/\xi-\log(x)/2$. This in general provides a more accurate estimate of the correlation length at long length scale. Second, we define a reduced correlation function
\begin{equation}
  \tilde C_{i,j}=\frac{\sqrt{|i-j|}}{e^{-|i-j|/\xi+c}}C_{i,j} 
\end{equation}
and fit it with a cosine $\tilde C_{i,j}\approx a\cos(q|i-j|+\varphi_0)$ as shown in Fig.\ref{fig:OZfit_example}(b). The agreement between the DMRG data (blue dots) and the result of the fit (red line) is almost perfect. Since in the first step we did not fit the peaks, but average out the oscillations, the factor $a\neq1$. 

\begin{figure}[h!]
\centering 
\includegraphics[width=0.49\textwidth]{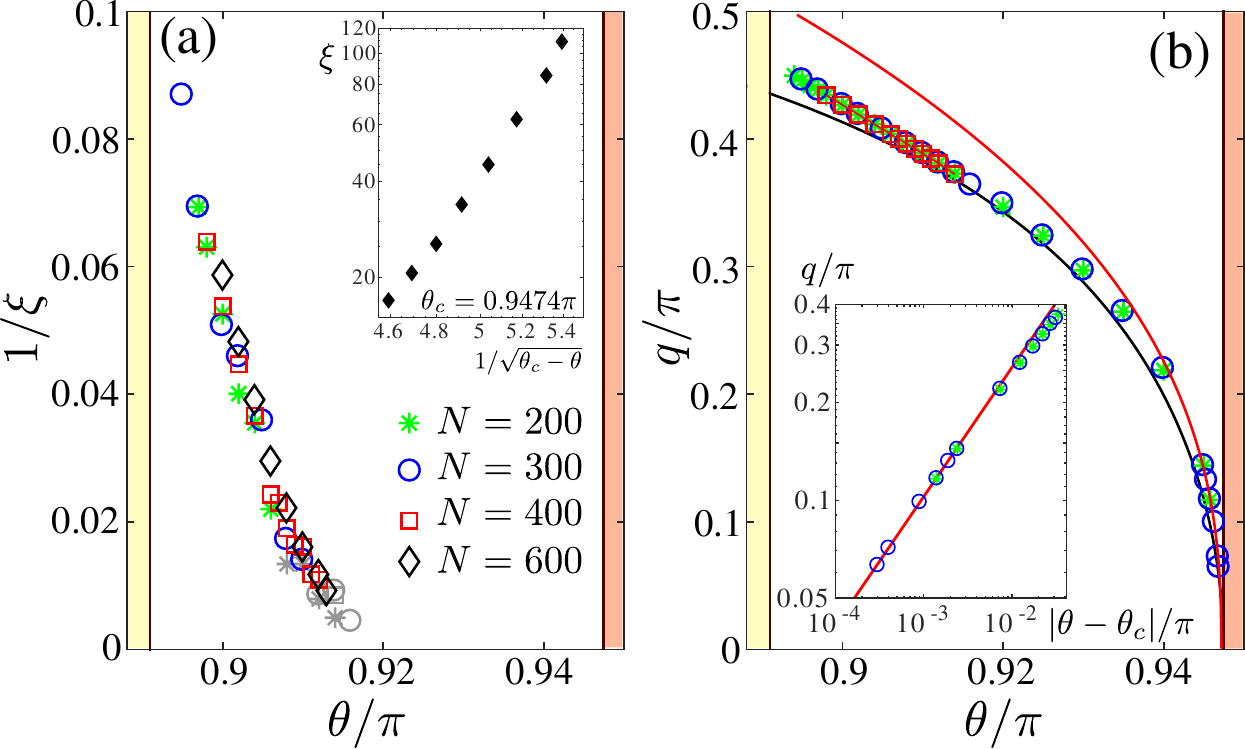}
\caption{(a) Inverse of the correlation length and (b) wave-vector $q$ as a function of $\theta$ between the dimerized and the ferromagnetic phases. Light-gray symbols stand for the points where the extracted correlation length is comparable to the total length of the chain. Inset : divergence of the correlation length with the distance to $\theta_3=0.9474\pi$ in a log-log scale. In panel (b), the black line corresponds to the $q$-vector given by Eq.\ref{eq:vave_vec_j1j3} that minimizes the energy of the one-magnon instability while the red line is the result of a fit $q\propto |\theta-\theta_c|^\nu$ with $\nu\approx0.40$. The inset shows the scaling towards the critical point in log-log scale. }
\label{fig:j1j3_correl_length}
\end{figure}

The results for the correlation length and the wave-vector as a function of $\theta$ are summarized in Fig.\ref{fig:j1j3_correl_length}.
The inverse of the correlation length decays very fast upon approaching the ferromagnetic phase with a concave shape typical of the exponential divergence of the correlation length at a Kosterlitz-Thouless transition. As shown in the inset of  Fig.\ref{fig:j1j3_correl_length} (a), assuming this to be the case is consistent with a Kosterlitz-Thouless phase transition at the value predicted by the one-magnon instability $\theta_3\approx 0.9474\pi$.  

The correlations remain incommensurate until the system enters the ferromagnetic phase. We extract the critical exponent $\nu$ assuming that the order parameter $q$ vanishes as $q\propto |\theta-\theta_c|^\nu$ upon approaching the critical line.  The value of the critical exponent  $\nu\approx0.40$.
The fit has been performed over the last six data points for $N=300$. 
The numerical results for $q$ agree remarkably well with the $q$ vector of the lowest-energy one-magnon state of Eq. \ref{eq:vave_vec_j1j3}, which is also plotted in Fig.\ref{fig:j1j3_correl_length} (b).


\section{Haldane chain with next-nearest-neighbor interaction}
\label{sec:j1j2}

\begin{figure}[t!]
\includegraphics[width=0.45\textwidth]{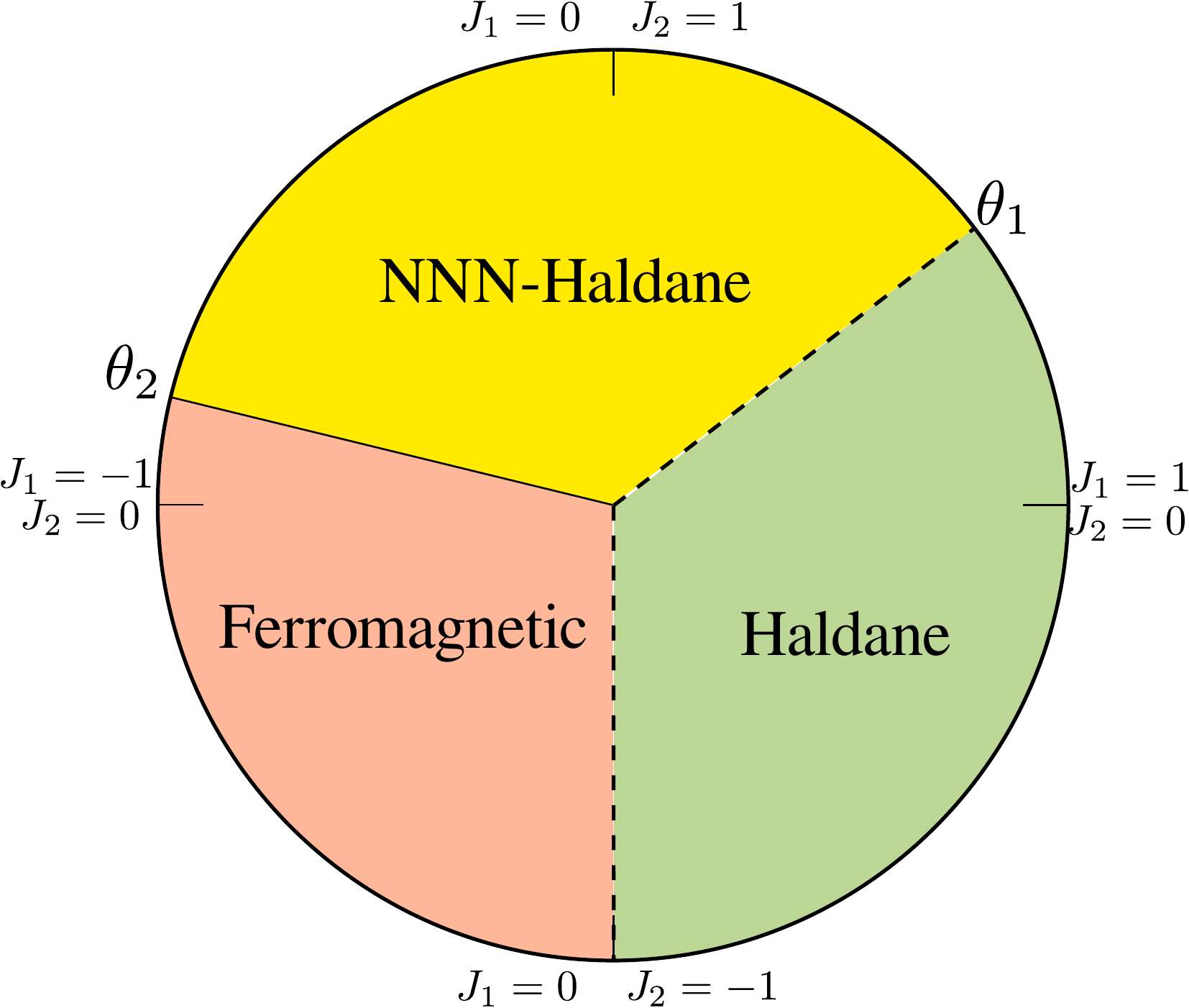}
\caption{Phase diagram of $J_1-J_2$ model given by the Hailtonian of Eq.\ref{eq:j1j2}. The transition between the Haldane and the NNN-Haldane phase is first order and takes place at $\theta_1\approx0.204\pi$. The system undergoes a continuous transition to the ferromagnetic phase at $\theta_2\approx0.9220\pi$. The transition between the ferromagnetic and the Haldane phase is first order and is at $\theta=3\pi/2$}
\label{fig:phase_diag_j1j2}
\end{figure}

In this section, we turn to the full phase diagram of the spin-1 $J_1-J_2$ model defined by the Hamiltonian of Eq.\ref{eq:j1j2}. 
It is summarized in Fig.\ref{fig:phase_diag_j1j2}. It contains three main phases - Haldane, NNN-Haldane and ferromagnetic. This phase diagram has been studied previously, and our results agree qualitatively with those of Ref.[\cite{lee}], but our study of the one-magnon instability and our DMRG results obtained on larger systems are useful to confirm the  nature of the phase transition between the ferromagnetic and the NNN-Haldane phases.

The transition between the Haldane and the NNN-Haldane phases takes place at $J_2/J_1\approx0.75$ ($\theta_1\approx 0.204\pi$) and it is a first order transition\cite{kolezhuk_prl}. On the other side of the Haldane phase the system undergoes a first-order transition into the ferromagnetic phase as soon as both couplings are ferromagnetic.

As for the $J_1-J_3$ model, we look at the one-magnon instability of the ferromagnetic phase. For the $J_1-J_2$ model, the energy of a single magnon state defined in Eq.\ref{eq:one_mag} takes the following form 
\begin{equation}
  E(q)=(J_1+J_2)(N-2)+2J_1\cos q+2J_2\cos 2q.
\end{equation}
Details of the derivation can be found in Appendix \ref{ap:one_magnon_j1j2}.
The curvature of the energy as a function of the wave-number $q$ is the given by:
\begin{equation}
  c=-2J_1\cos q-8J_2\cos 2q.
\end{equation}
As before, the transition is associated with the point where  the sign of the curvature computed at $q=0$ changes. This leads to $\frac{J_2}{J_1}=-\frac{1}{4}$, i.e. $\theta=0.9220\pi$, in good agreement with our DMRG data.
The energy of the single-magnon instability is minimized along the line 
\begin{equation}
  \cos q=-\frac{J_1}{4J_2}=-\frac{1}{4}\cot\theta
  \label{eq:j1j2_qmin}
\end{equation}

\begin{figure}[h!]
\centering 
\includegraphics[width=0.49\textwidth]{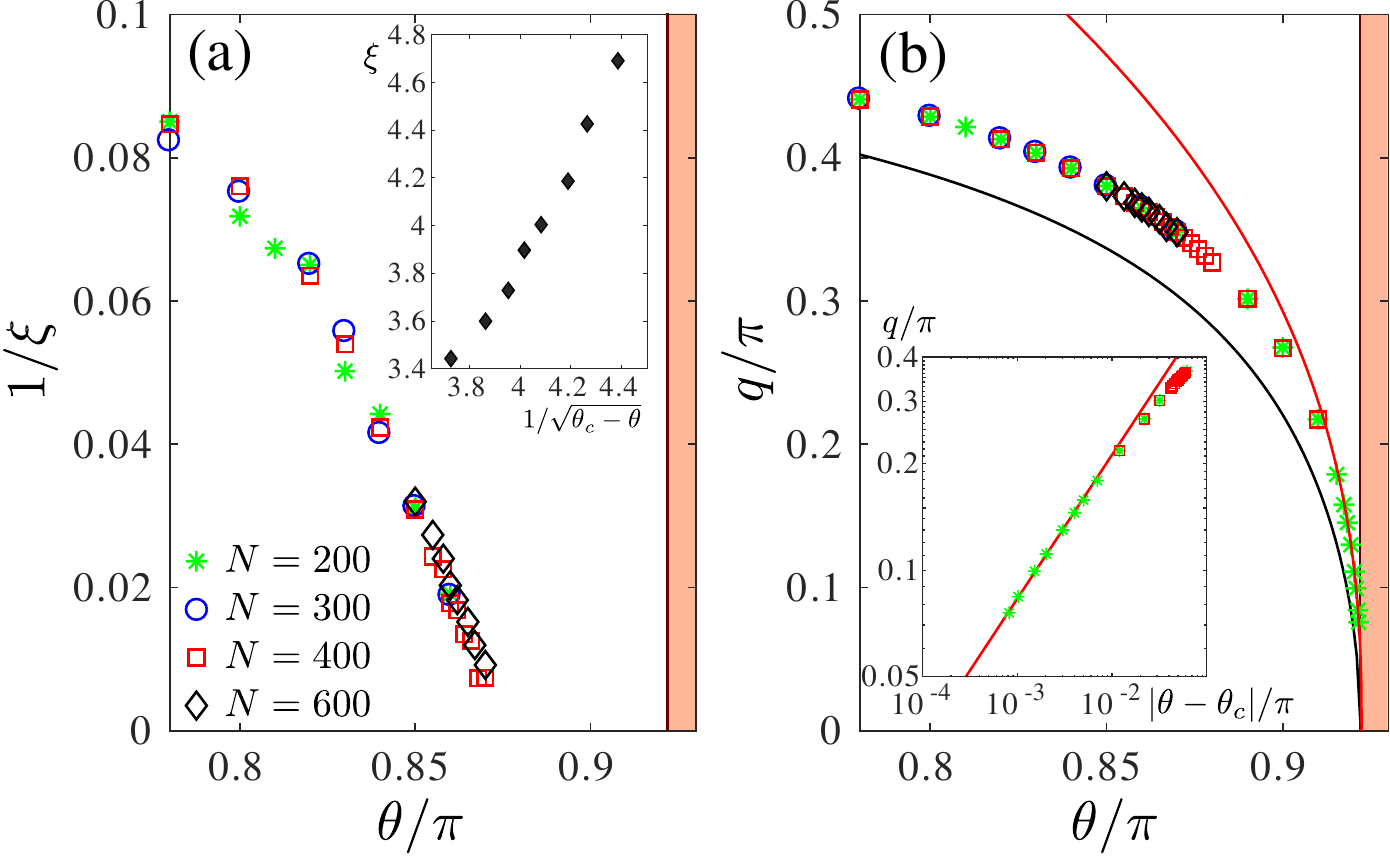}
\caption{(a) Inverse of the correlation length and (b) wave-vector $q$ as a function of $\theta$. In panel (b), the black line corresponds to the $q$-vector given by Eq.\ref{eq:j1j2_qmin} that minimizes the energy of the one-magnon instability while the red line is the result of a fit $q\propto |\theta-\theta_c|^\nu$ with $\nu\approx0.41$. The inset shows the scaling towards the critical point in log-log scale.}
\label{fig:j1j2_corlength}
\end{figure}

To study the transition between the NNN-Haldane and ferromagnetic phases, we have analyzed the NNN-Haldane phase along the same lines as for the $J_1-J_3$ model: We have fitted the spin-spin correlation to the Ornstein-Zernicke form in order to  extract the correlation length $\xi$ and the wave-vector $q$ as a function of $\theta$. The results obtained for various system sizes are presented in Fig.\ref{fig:j1j2_corlength}. As for the $J_1-J_3$ model we observe a fast decay of the inverse of the correlation length, as shown in Fig.\ref{fig:j1j2_corlength}(a). When plotted in a log-log scale the divergence of the correlation length as a function of the distance to the critical value is consistent with a Kosterlitz-Thouless transition at the critical point $\theta_2=0.922\pi$ where the one-magnon instability takes place. Note that in Ref.[\cite{lee}] the calculations have been performed on a chain with up to 200 sites keeping at most 300 states. In the present study we perform DMRG calculations on open chains with up to 600 sites, keeping up to 2000 states and making up to 10 DMRG sweeps. Close to the ferromagnetic phase the results we present here quantitatively disagree with the results from Ref.\onlinecite{lee}. For instance, at $\theta=0.87$ ($J_1/J_2\approx -2$) we obtained a correlation length of $\xi\approx 110$, while in Ref.\onlinecite{lee} the extrapolated value is only $\xi\approx 22$.

As shown in Fig.\ref{fig:j1j2_corlength}(b),the correlations remain incommensurate until the transition to the ferromagnetic phase at $\theta_2\approx0.9220\pi$. Interestingly enough the numerical data for $q$ as a function of $\theta$ seems to follow the line along which the energy for the single magnon instability is minimal, although the agreement is less spectacular than for $J_1-J_3$ model. By fitting the numerical data for $q$ we extract a critical exponent $\nu\approx0.41$.

\section{Discussion}
\label{sec:conc}

\begin{figure}[t!]
\includegraphics[width=0.45\textwidth]{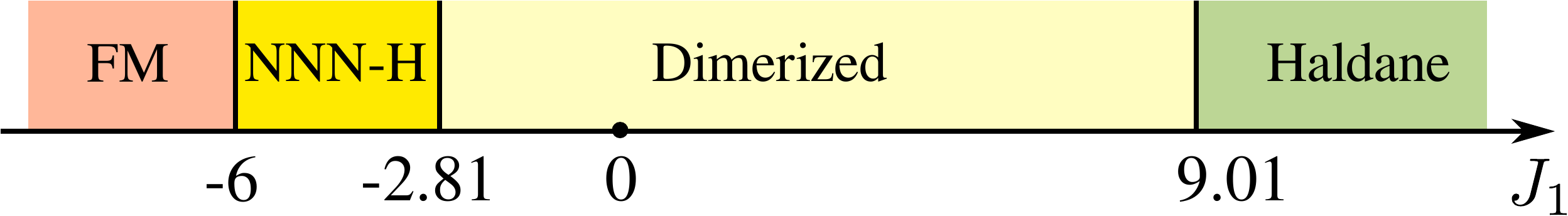}
\caption{Phase diagram as a function of nearest-neighbor coupling $J_1$ with fixed coupling constant $J_3=1$. }
\label{fig:pd_lin_j1j3}
\end{figure}

\begin{figure}[t!]
\includegraphics[width=0.45\textwidth]{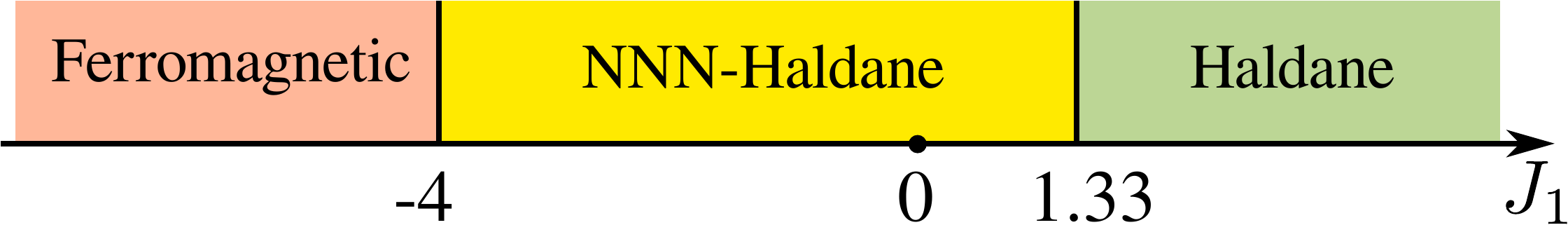}
\caption{Phase diagram as a function of nearest-neighbor coupling $J_1$ with fixed coupling constant $J_2=1$.}
\label{fig:pd_lin_j1j2}
\end{figure}

Using extensive DMRG calculations, we have determined the complete phase diagram of the spin-1  $J_1-J_2$ and $J_1-J_3$ models for both positive and negative signs of the coupling constants. While there is a direct, first-order transition between the Haldane phase and the ferromagnetic phase in both cases for negative $J_2$ or $J_3$, the phase diagrams are richer for positive couplings, and they are different as shown in Figs. \ref{fig:pd_lin_j1j3} and \ref{fig:pd_lin_j1j2}. The NNN-Haldane phase that has long been known for the $J_1-J_2$ model turns out to be also present in the $J_1-J_3$ model, but only close to the ferromagnetic phase. Between the NNN-Haldane phase and the Haldane phase, there is a large dimerized phase that extends deep into the negative $J_1$ region. So coupling spins 1 beyond nearest-neighbors leads generically to a physics which is not unlike that of the spin-1/2 case since there is the possibility of dimerization, but which is richer because this spontaneous dimerization has to compete with an effective decoupling between NNN spin-1 chains. The difference can be traced back to the nature of the ground state of the Heisenberg spin chains. For spin-1/2, the ground state has critical correlations, and NNN interactions immediately lead to dimerization. In the case of the spin-1 chain, the spectrum is gapped, and decoupled spin-1 chains are not immediately unstable towards dimerization. In the case of the $J_1-J_2$ model, the decoupling is expected to be effective close to $J_1=0$. It turns out that it remains present all the way to the ferromagnetic resp. to the Haldane phase. In the case of the $J_1-J_3$ chain, there is no limit where NNN chains are strictly decoupled, and the effective decoupling only shows up close to the ferromagnetic phase. 

Given the wealth of new phases induced by a magnetic field in the spin-1/2  $J_1-J_2$ chain\cite{sudan,furusaki} and the bilinear-biquadratic spin-1 chain\cite{manmana}, it will be interesting to study these models, in particular the $J_1-J_3$ model in which all phases are realized, in the presence of a magnetic field. This is left for future investigation.

\section{Acknowledgments}

This work has been supported by the Swiss National Science Foundation.
The calculations have been performed using the facilities of the Scientific IT and Application Support Center of EPFL and computing facilities of University of Amsterdam.

\bibliographystyle{apsrev4-1}
\bibliography{bibliography}

\begin{appendix}
  \section{One magnon instability from the ferromagnetic phase in the $J_1-J_3$ model}
  \label{ap:one_magnon_j1j3}

In the vicinity of the ferromagnetic phase, 
we start with the ferromagnetic wave-function $|0\rangle=|S,S,...,S,S\rangle$, such that $S_l^+|0\rangle=0$.

First, consider the action of the Hamiltonian of Eq. \ref{eq:j1j3} on the state $S^-_l|0\rangle$. Thte three nearest-neighbor terms give the following contribution:

\begin{align*}
a_1=J_1\frac{1}{2}\sum_{j=1}^N S^+_jS^-_{j+1}S^-_l|0\rangle=J_1\frac{1}{2}S^-_{l+1}S^+_{l}S^-_l|0\rangle
\\=\frac{1}{2}J_1S^-_{l+1}(2S^z_l+S^-_{l}S^+_l)|0\rangle \\=J_1S\cdot S^-_{l+1}|0\rangle
\end{align*}

\begin{align*}
a_2=J_1\frac{1}{2}\sum_{j=1}^N S^-_jS^+_{j+1}S^-_l|0\rangle=\frac{1}{2}J_1S^-_{l-1}S^+_{l}S^-_l|0\rangle 
\\=\frac{1}{2}J_1S^-_{l-1}(2S^z_l+S^-_{l}S^+_l)|0\rangle\\=J_1S\cdot S^-_{l-1}|0\rangle
\end{align*}

\begin{align*}
a_3=J_1\sum_{j=1}^N S^z_jS^z_{j+1}S^-_l|0\rangle\\
=J_1\sum_{j\neq l,l-1} S^z_jS^z_{j+1}S^-_l|0\rangle+2J_1S(S-1)S^-_l|0\rangle
\\=J_1((N-2)S^2+2S(S-1))\cdot S^-_l|0\rangle
\\
=J_1S(NS-2)\cdot S^-_l|0\rangle
\end{align*}

The contribution from the three-site terms are given by:

\begin{equation*}
b_{1,1}=J_3\frac{1}{4}\sum_{j=1}^N S^+_{j-1}S^-_{j}S^+_{j}S^-_{j+1}S^-_l|0\rangle=0
\end{equation*}
\begin{equation*}
c_{1,1}=J_3\frac{1}{4}\sum_{j=1}^N S^+_{j+1}S^-_{j}S^+_{j}S^-_{j-1}S^-_l|0\rangle=0
\end{equation*}
\begin{equation*}
b_{1,2}=J_3\frac{1}{4}\sum_{j=1}^N S^+_{j-1}S^-_{j}S^-_{j}S^+_{j+1}S^-_l|0\rangle=0
\end{equation*}
\begin{equation*}
c_{1,2}=J_3\frac{1}{4}\sum_{j=1}^N S^+_{j+1}S^-_{j}S^-_{j}S^+_{j-1}S^-_l|0\rangle=0
\end{equation*}

\begin{align*}
b_{1,3}=J_3\frac{1}{2}\sum_{j=1}^N S^+_{j-1}S^-_{j}S^z_{j}S^z_{j+1}S^-_l|0\rangle\\
=\frac{1}{2}J_3S^2\cdot S^-_{l+1}S^+_{l}S^-_l|0\rangle\\
=J_3\frac{1}{2}S^2\cdot S^-_{l+1}(2S^z_l+S^-_{l}S^+_l)|0\rangle\\
=J_3S^3\cdot S^-_{l+1}|0\rangle
\end{align*}

\begin{align*}
c_{1,3}=J_3\frac{1}{2}\sum_{j=1}^N S^+_{j+1}S^-_{j}S^z_{j}S^z_{j-1}S^-_l|0\rangle\\
=\frac{1}{2}J_3S^2\cdot S^-_{l-1}S^+_{l}S^-_l|0\rangle \\
=J_3S^2\cdot S^-_{l-1}(2S^z_l+S^-_{l}S^+_l)|0\rangle\\
=J_3S^3\cdot S^-_{l-1}|0\rangle
\end{align*}

\begin{equation*}
b_{2,1}=J_3\frac{1}{2}\frac{1}{4}\sum_{j=1}^N S^-_{j-1}S^+_{j}S^+_{j}S^-_{j+1}S^-_l|0\rangle=0
\end{equation*}
\begin{equation*}
c_{2,1}=J_3\frac{1}{4}\sum_{j=1}^N S^-_{j+1}S^+_{j}S^+_{j}S^-_{j-1}S^-_l|0\rangle=0
\end{equation*}

\begin{align*}
b_{2,2}=J_3\frac{1}{4}\sum_{j=1}^N S^-_{j-1}S^+_{j}S^-_{j}S^+_{j+1}S^-_l|0\rangle\\
=\frac{1}{4}J_3S^-_{l-2}S^+_{l-1}S^-_{l-1}S^+_{l}S^-_l|0\rangle\\
=\frac{1}{4}J_3 S^-_{l-2}(2S^z_{l-1}+S^-_{l-1}S^+_{l-1})(2S^z_{l}+S^-_{l}S^+_{l})|0\rangle\\
=J_3S^2\cdot S^-_{l-2}|0\rangle
\end{align*}

\begin{align*}
c_{2,2}=J_3\frac{1}{4}\sum_{j=1}^N S^-_{j+1}S^+_{j}S^-_{j}S^+_{j-1}S^-_l|0\rangle\\
=\frac{1}{4}J_3S^-_{l+2}S^+_{l+1}S^-_{l+1}S^+_{l}S^-_l|0\rangle \\
=\frac{1}{4}J_3 S^-_{l+2}(2S^z_{l+1}+S^-_{l+1}S^+_{l=1})(2S^z_{l}+S^-_{l}S^+_{l})|0\rangle\\
=J_3S^2\cdot S^-_{l+2}|0\rangle
\end{align*}

\begin{align*}
b_{2,3}=J_3\frac{1}{2}\sum_{j=1}^N S^-_{j-1}S^+_{j}S^z_{j}S^z_{j+1}S^-_l|0\rangle\\
=\frac{1}{2}J_3S^-_{l-1}S^+_{l}S^z_{l}S^z_{l+1}S^-_l|0\rangle \\
=\frac{1}{2}J_3S(S-1)\cdot S^-_{l-1}(2S^z_{l}+S^-_{l}S^+_{l})|0\rangle\\
=J_3S^2(S-1)\cdot S^-_{l-1}|0\rangle
\end{align*}

\begin{align*}
c_{2,3}=J_3\frac{1}{2}\sum_{j=1}^N S^-_{j+1}S^+_{j}S^z_{j}S^z_{j-1}S^-_l|0\rangle\\
=\frac{1}{2}J_3S^-_{l+1}S^+_{l}S^z_{l}S^z_{l-1}S^-_l|0\rangle \\
 =J_3\frac{1}{2}S(S-1)\cdot S^-_{l+1}(2S^z_{l}+S^-_{l}S^+_{l})|0\rangle\\
=J_3S^2(S-1)\cdot S^-_{l+1}|0\rangle
\end{align*}

\begin{align*}
b_{3,1}=J_3\frac{1}{2}\sum_{j=1}^N S^z_{j-1}S^z_{j}S^+_{j}S^-_{j+1}S^-_l|0\rangle\\
=J_3\frac{1}{2}S^z_{l-1}S^z_{l}S^+_{l}S^-_{l+1}S^-_l|0\rangle \\
=J_3\frac{1}{2} S^z_{l-1}S^z_{l}S^-_{l+1}(2S^z_{l}+S^-_{l}S^+_{l})|0\rangle\\
=J_3S^3\cdot S^-_{l+1}|0\rangle
\end{align*}

\begin{align*}
c_{3,1}=J_3\frac{1}{2}\sum_{j=1}^N S^z_{j+1}S^z_{j}S^+_{j}S^-_{j-1}S^-_l|0\rangle\\
=\frac{1}{2} J_3S^z_{l+1}S^z_{l}S^+_{l}S^-_{l-1}S^-_l|0\rangle\\
=\frac{1}{2} J_3S^z_{l+1}S^z_{l}S^-_{l-1}(2S^z_{l}+S^-_{l}S^+_{l})|0\rangle\\
=J_3S^3\cdot S^-_{l-1}|0\rangle
\end{align*}

\begin{align*}
b_{3,2}=J_3\frac{1}{2}\sum_{j=1}^N S^z_{j-1}S^z_{j}S^-_{j}S^+_{j+1}S^-_l|0\rangle\\
=J_3\frac{1}{2}S^z_{l-2}S^z_{l-1}S^-_{l-1}S^+_{l}S^-_l|0\rangle\\
= J_3\frac{1}{2} S^z_{l-2}S^z_{l-1}S^-_{l-1}(2S^z_{l}+S^-_{l}S^+_{l})|0\rangle\\
=J_3S^2(S-1)\cdot S^-_{l-1}|0\rangle
\end{align*}

\begin{align*}
c_{3,2}=J_3\frac{1}{2}\sum_{j=1}^N S^z_{j+1}S^z_{j}S^-_{j}S^+_{j-1}S^-_l|0\rangle\\
=J_3\frac{1}{2}S^z_{l+2}S^z_{l+1}S^-_{l+1}S^+_{l}S^-_l|0\rangle \\
=J_3\frac{1}{2} S^z_{l+2}S^z_{l+1}S^-_{l+1}(2S^z_{l}+S^-_{l}S^+_{l})|0\rangle\\
=J_3S^2(S-1)\cdot S^-_{l+1}|0\rangle
\end{align*}

\begin{align*}
b_{3,3}+c_{3,3}=2J_3\sum_{j=1}^N S^z_{j-1}S^z_{j}S^z_{j}S^z_{j+1}S^-_l|0\rangle\\
=2J_3\sum_{j\neq l-1,l,l+1}S^4S^-_l|0\rangle+4J_3S^3(S-1) S^-_l|0\rangle \\+2J_3S^2(S-1)^2 S^-_l|0\rangle \\
=2J_3S^2(2S(S-1)+(S-1)^2+(N-3)S^2) S^-_l|0\rangle\\
=2J_3S^2(NS^2+1) S^-_l|0\rangle
\end{align*}

By summing all the contributions of all the terms, we get:

\begin{align*}
HS^-_l|0\rangle
=\left(J_1S(NS-2)+2J_3S^2(NS^2+1) \right)S^-_l|0\rangle\\
+S\left(J_1+4J_3S^2-2J_3S\right)(S^-_{l-1}+S^-_{l+1})|0\rangle\\
+\left(J_3S^2 \right)(S^-_{l-2}+S^-_{l+2})|0\rangle
\end{align*}

Introducing the short-hand notation
\begin{align}
f_1=J_1S(NS-2)+2J_3S^2(NS^2+1)\\
f_2=S(J_1+4J_3S^2-2J_3S)\\
f_3=J_3S^2,
\end{align}
the action of the Hamiltonian on the one-magnon state takes the simple form:
\begin{multline}
  HS^-_l|0\rangle=f_1S^-_l|0\rangle+f_2(S^-_{l-1}+S^-_{l+1})|0\rangle\\+f_3(S^-_{l-2}+S^-_{l+2})|0\rangle
\end{multline}

For the case $S=1$, the coefficients become:

\begin{align}
f_1=E_\mathrm{FM}+2(J_3-J_1)\\
f_2=J_1+2J_3\\
f_3=J_3
\end{align}
where $E_\mathrm{FM}=J_1NS^2+2J_3NS^4$ is the energy of the ferromagnetic state. The state $S^-_l|0\rangle$  itself is not a ground state of the  Hamiltonian of Eq.\ref{eq:j1j3}. However, the following combinations of these states are eigenstates of the Hamiltonian of Eq.\ref{eq:j1j3}:

\begin{equation}
|\tilde{\varphi}(q)\rangle=\sum_{l=1}^Ne^{iql}S^-_l|0\rangle
\end{equation}
These states are not normalized yet. Their norm is given by:
\begin{align*}
\langle\tilde{\varphi}|\tilde{\varphi}\rangle=\sum_{l,k=1}^N\langle0|e^{iq(l-k)}S^+_kS^-_l|0\rangle\\
=\langle0|2S^z_l+S^-_lS^+_l|0\rangle \\
=\sum_{l=1}^N\langle0|2S^z_l|0\rangle=2SN
\end{align*}

Then the normalized one-magnon states are given by:
\begin{equation}
|\varphi(q)\rangle=\frac{1}{\sqrt{2NS}}\sum_{l=1}^Ne^{iql}S^-_l|0\rangle
\end{equation}

The energy of these states is given by

\begin{equation}
E(q)=f_1+2f_2\cos q+2f_3\cos 2q
\end{equation}

  \section{One magnon instability in $J_1-J_2$ model}
  \label{ap:one_magnon_j1j2}
The derivation of the energy of a single-magnon state for the $J_1-J_2$ model is rather simple. The nearest-neighbor contributions have been already computed in Appendix \ref{ap:one_magnon_j1j3}. The next-nearest-neighbor contributions can be obtained in a similar way:

\begin{align*}
d_1=J_2\frac{1}{2}\sum_{j=1}^N S^+_jS^-_{j+2}S^-_l|0\rangle=J_2\frac{1}{2}S^-_{l+2}S^+_{l}S^-_l|0\rangle
\\=\frac{1}{2}J_2S^-_{l+2}(2S^z_l+S^-_{l}S^+_l)|0\rangle \\
=J_2S\cdot S^-_{l+2}|0\rangle
\end{align*}

\begin{align*}
d_2=J_2\frac{1}{2}\sum_{j=1}^N S^-_jS^+_{j+2}S^-_l|0\rangle=\frac{1}{2}J_2S^-_{l-2}S^+_{l}S^-_l|0\rangle 
\\=\frac{1}{2}J_2S^-_{l-2}(2S^z_l+S^-_{l}S^+_l)|0\rangle\\=J_2S\cdot S^-_{l-2}|0\rangle
\end{align*}

\begin{align*}
d_3=J_2\sum_{j=1}^N S^z_jS^z_{j+2}S^-_l|0\rangle\\
=J_2\sum_{j\neq l,l-2} S^z_jS^z_{j+2}S^-_l|0\rangle+2J_2S(S-1)S^-_l|0\rangle
\\=J_2\left[(N-2)S^2+2S(S-1)\right]\cdot S^-_l|0\rangle
\\
=J_2S(NS-2)\cdot S^-_l|0\rangle
\end{align*}

By summing the contributions of all the terms, we get:

\begin{align*}
HS^-_l|0\rangle
=\left(J_1+J_2\right)S(NS-2)S^-_l|0\rangle\\
+J_1S(S^-_{l-1}+S^-_{l+1})|0\rangle
+J_2S(S^-_{l-2}+S^-_{l+2})|0\rangle\\
\end{align*}

  \end{appendix}

\end{document}